\documentclass[twocolumn]{aa} 

\usepackage{natbib}
\usepackage{amsmath}
\usepackage{graphicx}
\usepackage{txfonts}
\usepackage{textcomp}

\makeatletter
\def\@biblabel#1{}
\makeatother

\newcommand{\mjup}{M$_{\rm J}$\,}
\newcommand{\msun}{M$_\odot$\,}

\newcommand{\ms}{m\,s$^{-1}$\,}
\newcommand{\teff}{T$_{{\rm eff}}$\,}
\renewcommand{\cite}{\citealp}

\begin{document}




\title{The properties of planets around giant stars.\thanks{Based on observations collected at La Silla - Paranal Observatory under
programs ID's 085.C-0557, 087.C.0476, 089.C-0524 and 090.C-0345.}}

   \titlerunning{The properties of planets around giant stars}
   \author{M. I. Jones \inst{1,2}
           \and J. S. Jenkins \inst{2}
           \and P. Bluhm \inst{3}
           \and P. Rojo \inst{2,4}
           \and C. H. F. Melo \inst{5}} 

         \institute{Department of Electrical Engineering and Center of Astro-Engineering UC, Pontificia Universidad 
         Cat\'olica de Chile, Av. Vicuña Mackenna 4860, 782-0436 Macul, Santiago, Chile \\\email{mjones@aiuc.puc.cl}
         \and Departamento de Astronom\'ia, Universidad de Chile, Camino El Observatorio 1515, Las Condes, Santiago, Chile 
         \and Departamento de Astronom\'ia, Universidad de Concepción, Casilla 160-C, Concepción, Chile 
         \and Department of Astronomy, Cornell University
         \and European Southern Observatory, Casilla 19001, Santiago, Chile}

   \date{}

 
  \abstract
{More than 50 exoplanets have been found around giant stars, revealing different properties when compared to planets
orbiting solar-type stars. In particular, they are Super-Jupiters and are not found orbiting interior to $\sim$ 0.5 AU.} 
{We are conducting a radial velocity study of a sample of 166 giant stars aimed at studying the population of close-in planets
orbiting giant stars and how their orbital and physical properties are influenced by the post-MS evolution of the host star.}
{We have collected multi epochs spectra for all of the targets in our sample. 
We have computed precision radial velocities from FECH/CHIRON and FEROS spectra, using the I$_2$ cell technique and the simultaneous
calibration method, respectively. }
{We present the discovery of a massive planet around the giant star HIP\,105854.
The best Keplerian fit to the data leads to an orbital distance of 0.81 $\pm$ 0.03 AU, an eccentricity of 0.02 $\pm$ 0.03 and a projected
mass of 8.2 $\pm$ 0.2 \mjup. 
With the addition of this new planet discovery, we performed a detailed analysis of the orbital properties 
and mass distribution of the planets orbiting giant stars.
We show that there is an overabundance of planets around giant stars with $a \sim$ 0.5-0.9 AU, which might be attributed to tidal decay.
Additionally, these planets are significantly more massive than those around MS and subgiant stars, suggesting that they grow via
accretion either from the stellar wind or by mass transfer from the host star. Finally, we show that planets around evolved
stars have lower orbital eccentricities than those orbiting solar-type stars, which suggests that they are either formed in different
conditions or that their orbits are efficiently circularized by interactions with the host star.  }
   {}


   \maketitle
%

\section{Introduction}

For the past two decades, different methods have been developed to detect and characterize extrasolar planets, giving rise to 
the discovery of more than 1000 planetary systems. Among them, the radial velocity (RV) technique has led
to the detection of a significant fraction of these systems. Unfortunately, this method is strongly sensitive to the spectral properties of the 
object, and hence it is mainly restricted to solar-type stars. On the one hand, very low-mass stars (such as mid and late M stars) 
are faint and their spectral distribution peaks in the infrared (IR) region, hence they have been mostly excluded in optical RV surveys. 
On the other hand, intermediate-mass and high-mass stars (M$_\star$ $\gtrsim$ 1.5 \msun) are too hot and thus they have an optical spectrum that is characterized by few and broad lines.
However, after the main-sequence (MS), they expand becoming cooler, and also their rotational velocity is strongly reduced. As a result, 
their spectra present thousands of narrow absorption lines, becoming suitable targets for precision RV surveys. 
Therefore, enough RV information can be obtained to search for planets around intermediate-mass stars and to 
study the dynamical interaction between them and the host star during the giant phase.
Unfortunately, giant stars present a high level of jitter (e.g. Sato et al. \cite{SAT05}), 
limiting the detection only to giant planets (K$\gtrsim$ 30 \ms). \newline \indent
So far, more than 50 giant planets have been found around giant stars, revealing interesting properties that seem to contrast 
the properties of giant planets discovered around dwarf stars. 
First, there is no planet orbiting any giant star in orbits interior
to $a$ $\sim$ 0.5 AU, in contrast to what is observed in solar-type stars, where there is a large number of discovered close-in gas 
giants.
This observational result suggests that the innermost planets are destroyed by the host star during the expansion phase of the stellar 
envelope. Moreover,
several theoretical studies have shown that the tidal interaction between the planet and the stellar convective envelope leads to 
the loss of orbital angular momentum. As a result, close-in planets are expected to spiral inward and hence are subsequently engulfed 
by the host star (e.g. Livio \& Soker \cite{LIV83}; Kunitomo et al. \cite{KUN11}). However, Johnson et al. (\cite{JOH07}) showed that 
close-in planets are also absent around intermediate-mass subgiant stars. Additionally, Bowler et al. \cite{BOW10} confirmed the 
Johnson et al. findings and also they showed that planets around intermediate-mass stars tend to reside at a larger orbital distance ($a$ 
$\gtrsim$ 1 AU) 
than planets around solar-type stars. According to these results, planets around giant stars are not found in close-in orbits because
of a different formation scenario around intermediate-mass stars instead of a dynamical evolutionary process, such as tidal
orbital decay. \newline \indent
Second, the mass distribution of planets around giant stars is different than for planets around dwarf stars. 
In particular, their occurrence rate is higher and they are on average more massive (D\"{o}llinger et al. \cite{DOL09}). 
This observational result is supported by theoretical studies that predict an increase in the frequency of giant planets with the 
mass of the host star (e.g. Kennedy \& Kenyon \cite{KEN08}).  \newline \indent
Last, planet-hosting giant stars are on average metal-poor, in direct contrast to the well known planet-metallicity connection that
is observed in MS stars (Gonzalez \cite{GON97}; Santos et al. \cite{SAN01}; Fischer \& Valenti \cite{FIS05}), which seems to be also 
present in the low-mass planetary regime (Jenkins et al. \cite{JEN13a}). 
Following the discussion presented in Schuler et al. (\cite{SCH05}), Pasquini et al. (\cite{PAS07}) argued that the enhanced metallicity of MS host stars is explained by pollution, instead 
of being a primordial feature of the protoplanetary disk. In this scenario, a significant amount of material, including planetesimals
and planet cores, fall into the star, hence increasing the metal abundance in its atmosphere. However, after these polluted stars 
evolve off the main-sequence, they develop large convective envelopes, where the metal excess is diluted. This idea might explain 
the aforementioned discrepancy between MS and giant host stars metallicity distribution. However, more recently, Mortier et al. 
(\cite{MOR13}) showed that RV surveys of giant stars are biased toward metal-poor stars, which explains why planets are 
preferentially found around metal-poor giants. On the other hand, giant planets around intermediate-mass stars might be efficiently
formed by the disk instability mechanism (Boss 1997). In this scenario the metal content of the protoplanetary disk does not play
a significant role on the formation of planets, hence it might explain why we do not see a trend toward metal richness in
host giant stars. \newline \indent
In this paper, we present the discovery of a massive planet around the red giant branch (RGB) star HIP\,105854, as part of the 
EXPRESS (EXoPlanets aRound Evolved StarS) project (Jones et al.  \cite{JON11,JON13}) and we present a detailed analysis of the 
properties of exoplanets around post-MS stars. 
In $\S$ 2 and $\S$ 3, we describe the observations, the data reduction and the RV computation method. 
In $\S$ 4 and $\S$ 5, we discuss the stellar properties of
the host star and the derived orbital parameters of HIP\,105854\,b. 
In $\S$ 6, we present the line profile analysis and the study of the 
photometric variability of HIP\,105854. In $\S$ 7, we summarize the properties of the planets detected around
giant stars and we discuss the effect of the stellar evolution on these properties. Finally, the conclusions are 
presented in $\S$ 8. 

\section{A massive planet around the intermediate-mass giant star HIP\,105854}

\subsection{Observations and data reduction}

A total of 13 spectra of HIP105854 were taken on seven different nights using the Fiber-fed Extended Range Optical Spectrograph (FEROS; Kaufer et al. 
\cite{KAU99}), on the 2.2m telescope, at La Silla Observatory. FEROS has a mean resolving power of R $\sim$ 48'000, covering
the entire optical spectral range. The typical observing time was 70-150 seconds (depending on the 
atmospheric conditions), giving rise to a signal-to-noise ratio (S/N) of $\sim$ 100-200, at 5500 \AA.
The spectra were extracted with the FEROS Data Reduction System, which performs a bias subtraction,
the flat-field correction, and order extraction. The wavelength solution was computed  
from a total of $\sim$ 1000 emission lines, which are identified in two different ThArNe lamps
that were taken either at the beginning of the night, or in the morning, just after the end of the observations.
The typical RMS in the wavelength calibration is $\sim$ 0.005 \AA.
The barycentric correction performed by the FEROS pipeline was disabled. \newline \indent
Additionally, we obtained 36 spectra using the old fiber echelle spectrograph (FECH) and CHIRON (Tokovinin et al. \cite{TOK13}), 
both of them installed in the 1.5m 
telescope, at The Cerro Tololo Inter-American Observatory. The two spectrographs are equipped with an I$_2$ cell that is placed in the light beam, which 
superimposes thousands of narrow absorption lines in the spectral region between $\sim$ 5000-6000 \AA. 
The I$_2$ absorption spectrum was used as a precise wavelength reference (see section \ref{sec_RVC}). 
We used the narrow slit with FECH (R $\sim$ 45'000) and the slit mode with CHIRON (R $\sim$ 90'000). The typical observing 
time was 700-800 seconds, leading to a 
S/N $\gtrsim$ 50. The extraction of the data was done in a similar fashion as that for the FEROS spectra, using the automatic 
reduction pipeline that is offered to CHIRON users. For more details see Jones (\cite{JONES13}). 

\begin{figure}[ht]
\centering
\includegraphics[width=7cm,height=9cm,angle=270]{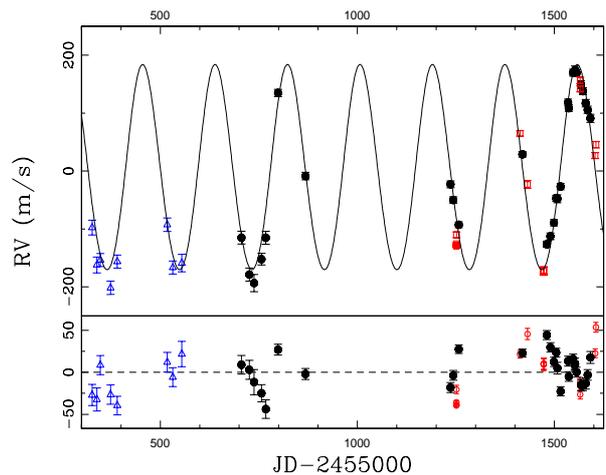}
\caption{Upper panel: Radial velocity curve for HIP\,105854. The open blue triangles, black filled dots and red open squares 
correspond to FECH, CHIRON and FEROS velocities, respectively. The best single-planet solution is overplotted (black solid line). 
Lower panel: Residuals from the Keplerian fit. The RMS is 23.6 \ms.  \label{HIP105854_vels}}
\end{figure}
\begin{figure}[ht]
\centering
\includegraphics[width=9cm,height=7cm]{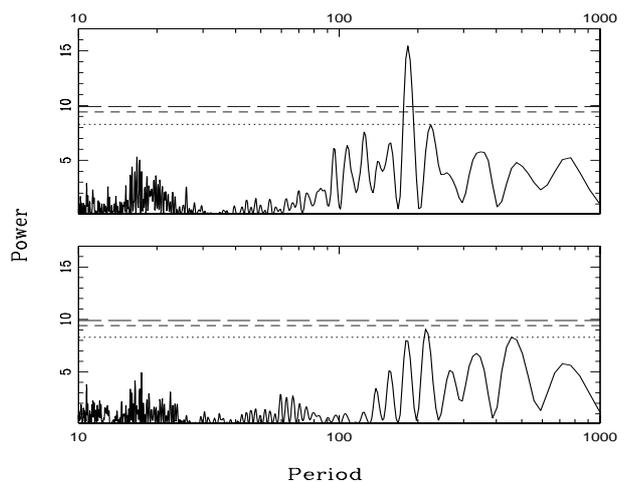}
\caption{LS periodogram of the HIP105854 radial velocities (upper panel) and the corresponding window function 
(lower panel). The three lines from bottom to top correspond to 1\%, 0.1\% and 0.01\% false alarm probability, respectively.
\label{HIP105854_periodogram}}
\end{figure}

\subsection{Radial velocity computations \label{sec_RVC}}

The radial velocities (RVs) from the FEROS spectra were computed by applying the simultaneous calibration method
(Queloz et al. \cite{QUE99}). The details of the calculations are explained in Jones et al. (\cite{JON13}) and 
Jones \& Jenkins (\cite{JONJEN14}), but the basic ideas of the procedure are also described below. \newline \indent
First, we computed the cross-correlation function (CCF) between the observed spectrum 
and a template. In this case the template corresponds to a high S/N observation of HIP\,105854.
We repeated this procedure on chunks of $\sim$ 50 \AA \,width (four chunks per order, from 35 different orders). 
We then used an iterative rejection code to compute the mean radial velocity from all of the chunk velocities. 
In a similar fashion, we also applied this procedure to the sky fiber, which is illuminated during the science observation
by a ThArNe lamp. This method allows us 
to compute the spectral drift, which is added to the RV measured with respect to the stellar template.
Finally, the barycentric velocity is computed and is added to the stellar velocity. 
It is worth noting that the velocity drift and the barycentric velocity can take either negative or positive values.
The typical single shot precision achieved for this star using our code is $\sim$ 5 \ms.  \newline  \indent
The FECH and CHIRON velocities were computed using the I$_2$ cell technique (Butler et al. \cite{BUT96}). The basic idea of 
this method is to pass the stellar light through a cell that contains molecular iodine vapor, which superimposes thousands 
of absorption lines in the wavelength region between 5000 \AA\, and 6000 \AA. These absorption lines are used as
wavelength markers against which the spectral doppler shift can be measured. We followed a similar procedure as the one 
described in Butler et al. (\cite{BUT96}), although with some small modifications. For instance, we computed the RVs on 
chunks of 5 \AA\,  (instead of 2 \AA) and we used a simpler model for the instrumental Point Spread Function (PSF), including only three Gaussians. 
We found that this combination of chunk size and PSF modeling yields to the best RV precision, especially for the FECH spectra. 
We obtained a typical radial velocity precision of $\sim$ 11 \ms for FECH data and $\sim$ 6 \ms using CHIRON spectra. 
Further details can be found in Jones (\cite{JONES13}). 
\begin{table}
\centering
\caption{Stellar parameters of HIP\,105854 \label{stellar_par}}
\begin{tabular}{lcc}
\hline\hline
Parameter      & Value \\
\hline
Spectral Type         &  K2III \\
B-V (mag)             &  1.2   \\
V (mag)               &  5.64  \\ 
Parallax (mas)        &  12.37 $\pm$ 0.31 \\
Distance (pc)         & 80.84 $\pm$  2.03 \\ 
\teff (K)             & 4780 $\pm$ 100 \\
log\,L (L$_\odot$)    & 1.670 $\pm$ 0.046 \\
log\,g (cm\,s$^{-1}$) & 2.94 $\pm$ 0.20 \\
{\rm [Fe/H]} (dex)    & 0.31 $\pm$ 0.18 \\
M$_\star$ (\msun)     & 2.1 $\pm$ 0.1 \\
\hline\hline
\end{tabular}
\end{table}

\subsection{Stellar properties}

HIP\,105854 is listed in the Hipparcos catalog as a K2III star, with B-V color equal to 1.20 and visual magnitude of 5.64
The parallax of 12.37 corresponds to a distance of 80.84 parsecs. Based on the equivalent width of more than 100 iron lines, and by
imposing local thermodynamic equilibrium, 
Jones et al. (\cite{JON11}) derived the following atmospheric parameters for HIP\,105854: Teff = 4780 $\pm$ 100 K, logL = 1.670 
$\pm$ 0.046 L$_\odot$, logg = 2.94 $\pm$ 0.20 cm\,s$^{-1}$ and [Fe/H] = 0.31 $\pm$ 0.18 dex. It is worth mentioning that 
the photometric and spectroscopic values of logg are systematically different (see discussion in Jones et al. \cite{JON11}).
By comparing these quantities with Salasnich et al. (\cite{SAL00}) evolutionary tracks, they identified this object as a 
post-RGB star with M$_\star$ = 2.1 $\pm$ 0.1 \msun. These values are summarized in Table \ref{stellar_par}.

\subsection{Orbital parameters of HIP\,105854\,b}

For the past three years, we have collected a total of 49 high resolution and high S/N spectra of the evolved star
HIP\,105854, using three different instruments, namely FECH, CHIRON and FEROS. From these datasets we measured precision 
RV variations, which have revealed a large amplitude periodic signal. 
Figure \ref{HIP105854_vels} shows the radial velocity curve of HIP\,105854. The blue open triangles and black filled dots 
correspond to FECH and CHIRON velocities, respectively, whereas the red open squares to FEROS RVs.
The FEROS and FECH/CHIRON radial velocity variations are listed in 
Tables \ref{feros_HIP105854_vels} and \ref{chiron_HIP105854_vels}, respectively\footnote{The RVs are relative to the mean values of the 
two datasets. There is also a zero point offset between the FEROS and FECH/CHIRON velocities}.
The solid line corresponds to the best Keplerian fit including only one planet. The solution was computed with the Systemic console 
(Meschiari et al.  \cite{MES09}). The uncertainties in the orbital parameters were estimated using the bootstrap randomization 
method.
The residuals of the fit are plotted in the lower panel. The RMS is 23.6 \ms, which is consistent with 
the amplitude of stellar oscillations observed in giant stars (e.g. Frink et al. \cite{FRI01}). However, it seems that 
there is also a long term linear trend in the residuals, suggesting the presence of a distant stellar companion. 
By including a linear trend, the best solution leads to a 
smaller mass for HIP\,105854\,b ($\sim$ 7.4 \mjup), while the orbital period and eccentricity remain unchanged.
%
The derived orbital parameters are listed in Table \ref{orbital_parameters}. The offset velocities of FECH/CHIRON and FEROS 
data are also listed ($\gamma_1$ and $\gamma_2$, respectively). 
Additionally, Figure \ref{HIP105854_periodogram} shows the Lomb-Scargle periodogram (Scargle 
\cite{SCA82}) of the HIP105854 RVs (upper panel). 
The three lines from bottom to top correspond to a false alarm probability of 1\%, 0.1\% and 0.01\%, respectively.
As can be seen there is a strong peak around 184 days. Also, the lower 
panel shows the periodogram of the window function. Clearly the peak observed in the RVs is not explained by a periodicity in the
sampling. \newline \indent
Since the period of the signal is close to half a year, we scrutinized
in great detail the calculation of the barycentric correction that could
lead to such an artifact.  We found no problem in the coordinates or
dates.  Moreover, we computed the barycentric correction using two
independent codes (based on FORTRAN and IDL, respectively), which
yielded velocities that agree within 1 \ms for this planet and have
found no such periods in more than the 100 stars we are monitoring.
\begin{table}
\centering
\caption{FEROS radial velocity variations of HIP\,105854\label{feros_HIP105854_vels}}
\begin{tabular}{lcc}
\hline\hline
JD\,-        & RV & error  \\
2455000   &  (\ms)   &   (\ms) \\
\hline
1251.540  &  -64.1  &5.0 \\
1251.597  &  -82.4  &1.7 \\
1251.600  &  -79.6  &1.9 \\
1251.602  &  -83.4  &1.6 \\
1412.805  &  111.0  &4.6 \\
1431.821  &   22.9  &6.7 \\
1472.822  & -126.1  &6.6 \\
1472.853  & -127.0  &7.2 \\
1472.890  & -126.1  &6.8 \\
1565.558  &  202.5  &5.7 \\
1565.615  &  187.8  &5.4 \\
1603.591  &   72.8  &5.2 \\
1605.623  &   91.6  &5.7 \\
\hline\hline
\end{tabular}
\end{table}
\begin{table}
\centering
\caption{FECH\,/\,CHIRON radial velocity variations of HIP\,105854\label{chiron_HIP105854_vels}}
\begin{tabular}{lcc}
\hline\hline
JD\,-        & RV & error  \\
2455000   &  (\ms)   &   (\ms) \\
\hline
   326.900 &  -68.1 & 12.6 \\
   338.851 & -133.8 & 13.7 \\
   347.865 & -125.4 & 11.3 \\
   373.746 & -173.0 & 11.5 \\
   390.762 & -129.5 & 11.1 \\
   517.551 &  -65.4 & 11.6 \\
   531.553 & -139.0 & 11.2 \\
   554.519 & -131.7 & 15.1 \\
   705.838 &  -86.0 & 11.0 \\
   725.866 & -150.4 & 11.2 \\
   737.903 & -163.9 & 14.9 \\
   756.807 & -125.2 & 10.1 \\
   767.798 &  -86.6 & 11.1 \\
   798.755 &  163.0 &  6.4 \\
   868.494 &   18.5 &  6.6 \\
  1236.531 &    5.1 &  5.9 \\
  1243.537 &  -21.8 &  5.7 \\
  1257.551 &  -64.8 &  5.1 \\
  1418.928 &   57.8 &  4.8 \\
  1480.818 &  -98.4 &  5.4 \\
  1489.911 &  -84.4 &  5.6 \\
  1498.898 &  -61.3 &  5.6 \\
  1504.699 &  -18.8 &  5.3 \\
  1507.754 &  -19.6 &  7.0 \\
  1515.805 &    1.2 &  5.4 \\
  1534.849 &  146.6 &  5.9 \\
  1536.687 &  136.2 &  5.7 \\
  1547.655 &  197.7 &  5.8 \\
  1551.601 &  199.0 &  5.1 \\
  1552.541 &  201.6 &  7.1 \\
  1556.527 &  196.8 &  5.9 \\
  1567.496 &  177.2 &  6.9 \\
  1572.499 &  166.4 &  5.9 \\
  1579.517 &  145.9 &  6.6 \\
  1584.530 &  134.3 &  6.3 \\
  1591.523 &  119.6 &  7.1 \\
\hline\hline
\end{tabular}
\end{table}
 
\begin{table}
\centering
\caption{Orbital parameters of HIP\,105854\,b \label{orbital_parameters}}
\begin{tabular}{l c}
\hline\hline
Parameter         &    Value     \\
\hline
P (days)          &   184.2 $\pm$ 0.5        \\
K (\ms)           &   178.1 $\pm$ 10.0       \\
a (AU)            &   0.81  $\pm$ 0.03       \\
e                 &   0.02  $\pm$ 0.03       \\
$\omega$ (deg)    &   343.2 $\pm$ 4.9        \\
M\,sin$i$ (\mjup) &   8.2   $\pm$ 0.2        \\
T$_0$ (JD)        &   2455262.4   $\pm$ 2.4  \\
$\gamma_1$ (\ms)  &   14.4  $\pm$ 4.2        \\
$\gamma_2$ (\ms)  &   22.3  $\pm$ 9.0        \\
\hline\hline
\end{tabular}
\end{table}

\subsection{Line profile analysis and stellar photometric variability \label{sec_BVS}}

\begin{figure}[h!]
\centering
\includegraphics[width=7cm,height=9cm,angle=270]{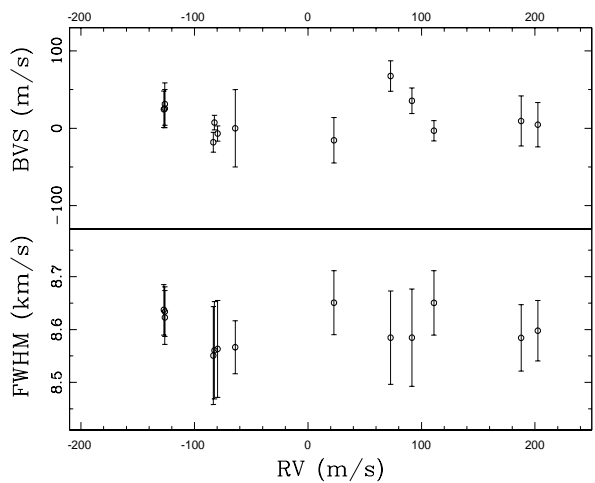}
\caption{Upper panel: Bisector velocity span against the FEROS RVs. The BVS at each epoch was computed from
the mean BVS in 11 different orders. The error bars correspond to the uncertainty in the mean.
Lower panel: FWHM of the CCF versus the measured RVs. In this case we computed the mean value from three consecutive orders.  \label{HIP105854_BVS}}
\end{figure}

\begin{figure}[h!]
\centering
\includegraphics[width=7cm,height=9cm,angle=270]{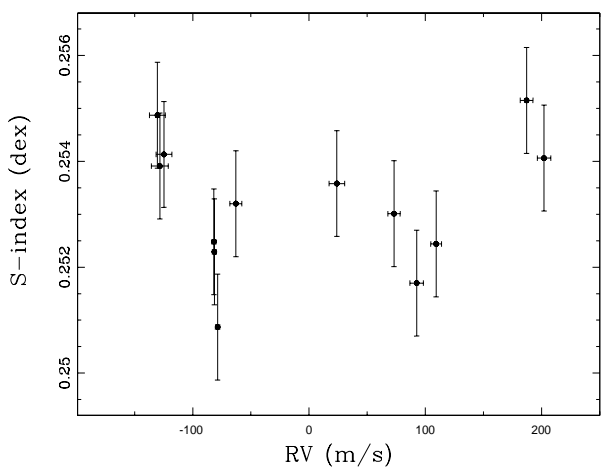}
\caption{Stellar chromospheric indicators (S-index) versus the measured radial velocities for HIP\,105854. The S-index are computed 
following the procedure described in Jenkins et al. (\cite{JEN08,JEN11}). \label{s_index}}
\end{figure}

To investigate the nature of the RV signal present in our data we have performed a line profile analysis and we studied the 
photometric variability of HIP\,105854. 
Figure \ref{HIP105854_BVS} shows the result of the line profile analysis. We included only FEROS data since the FECH and CHIRON 
spectra also contain the I$_2$ features. 
The bisector velocity span (BVS) and full width at half maximum (FWHM) of the CCF are plotted as as function of the measured radial velocities 
(upper and lower panel, respectively). 
The BVS values were computed for 11 different orders, giving 
rise to a mean BVS value at each observational epoch. Similarly, we computed the mean FWHM in three consecutive orders
($\Delta \lambda \sim 500 \AA$), since the width of the CCF is wavelength dependent. The error bars correspond to the uncertainty in 
the mean of these two quantities. 
As can be seen in Figure \ref{HIP105854_BVS}, our BVS and FWHM values are consistent with a flat line. 
This result shows that no mechanism causing line asymmetries seems to be responsible for the observed RV signal. \newline \indent
In addition, we computed stellar chromospheric activity, as described in Jenkins et al. (\cite{JEN08,JEN11}). The results are 
plotted in Figure \ref{s_index}. As can be seen, no obvious correlation is present between the S-index and the measured RVs. 
\newline \indent
Finally, to study whether the observed period is due to rotational modulation we measured the rotational period of the host star. 
Based on the derived stellar radius and projected rotational velocity (see Jones et al. \cite{JON11}) we 
obtain P$_{\rm R}$ = 137 $\pm$ 36 days. It can be noted that the orbital and rotational periods depart by more than one sigma.
Moreover, by assuming a mean inclination angle of $i$ = 45$\,^{\circ}$, the two values differ by more than 2 sigma. 
In addition, we analyzed the Hipparcos photometric data of HIP\,105854.
This dataset consists of 62 reliable H$_p$  observations 
(quality flag 0 or 1), taken between Julian dates 2447975.3 and 2448975.6.
The photometric variability is at the 0.006 mag level, and no significant periodicity is observed in the LS periodogram.
This level of variability in slowly rotating stars (such as HIP\,105854) is well below the predicted photometric variations
that are accompanied by a RV amplitude at the $\sim$ 200 \ms (Hatzes \cite{HAT02}). 
We can therefore discard the presence of spots (at least at the time that the Hipparcos data were obtained), hence it is unlikely that 
rotational modulation is the mechanism responsible for the periodic RV signal observed in the HIP\,105854 data.  


\section{The properties of planets around giant stars}

So far, more than 50 planets have been detected around giant stars, which have revealed
different properties than those orbiting solar-type stars. In this section, we review 
the semi-major axis, minimum mass and eccentricity distributions of these objects. Additionally, we discuss about the role of the 
stellar mass and stellar evolution on these properties.

\subsection{Orbital distribution}
After the detection of the first few planets around evolved stars, it was noticed that they are not found
in orbits interior to $\sim$ 0.7 AU (e.g. Johnson et al. \cite{JOH07}; Sato et al. \cite{SAT08}). 
Nowadays, more than 100 exoplanets have been discovered around such stars (including subgiants), and the paucity of short-period
systems is still present. Figure \ref{semiaxis_star} (left) shows the minimum mass of planets (up to 13 \mjup) 
as a function of the orbital distance, for planets around subgiant (blue triangles) and giant (red circles) stars,
respectively. For comparison, the position of the planets around solar-type stars (0.7 \msun > M$_\star$ > 1.2 \msun)
are overplotted (small black dots).
The dotted lines correspond to a radial velocity amplitude of 30 \ms (assuming M$_\star$ = 
1.5 \msun and $a$ = 1.0 AU, respectively), hence the planets above these lines should be easily detected by the current RV surveys.
It can be seen that there is an absence of planets orbiting giant stars at orbital distances below $\sim$ 0.5 AU
\footnote{HIP13044 b was shown not to be a planet (Jones \& Jenkins \cite{JONJEN14}).}
\footnote{There is a planet candidate in a 6.2-days orbit detected via light curve variations (Lillo-Box et al. \cite{LILLO14})}. 
This observational result has been attributed to be due to the dynamical interaction between the host 
stars and their planets. Different authors have studied this effect,
showing that close-in planets spiral inward due to the loss of orbital angular 
momentum, which is mainly transferred and dissipated in the stellar convective envelope (e.g. Livio \& Soker, \cite 
{LIV83}; Rasio et al. \cite{RAS96}; Sato et al. \cite{SAT08}; Schr\"oder \& Connon Smith \cite{SCH08}; 
Villaver \& Livio \cite{VIL09}; Kunitomo, et al. \cite{KUN11}). As a result of the tidal decay, close-in planets are 
not expected to be found, at least, around post-RGB stars. 
However, this theoretical prediction is far from being 
completely in agreement with the observed orbital properties of planets around giant stars. 
For instance, Villaver \& Livio 
(\cite{VIL09}) predict that the minimum orbital separation at which a planet survives around a 2 \msun star is
2.1 AU and 2.4 AU for planets with 1 \mjup and 3 \mjup, respectively, which does not agree with
the observational results, as shown in Figure \ref{semiaxis_star}. This is because most of the planet-hosting giants
are clump stars, thus most of them are core He-burning stars (i.e., post-RGB). However, based on a similar approach,
Kunitomo et al. (\cite{KUN11}) showed that the theoretical minimum survival distance for planets around giant stars, 
is in agreement with the observational results, also showing that the main differences with Villaver \& Livio 
predictions arise in the stellar evolutionary models. In fact, different evolutionary models predict very 
different stellar radii during the final stage of the RGB phase. This is certainly the largest 
uncertainty in the tidal decay term, which is strongly dependent on the ratio of the stellar radius and the orbital distance. \newline \indent
It can be also seen in 
Figure \ref{semiaxis_star} that planets around subgiants reside on average at larger orbital distances than those
orbiting giants stars, with the exception of the short period planets ($a$ $\lesssim$ 0.2 AU).
These close-in substellar objects are likely destroyed by the host star during the end of the RGB phase, when the stellar 
radius becomes larger than the orbital separation. 
In particular, there is an overabundance of planets around giant stars orbiting between $\sim$ 0.5 and 0.9 AU.
Indeed, there are only two planets around subgiant stars in this region, in contrast to 16 of them orbiting giant stars, corresponding to
$\sim$ 4\,\% and $\sim$ 32\,\% of the total population, respectively.
This result might suggest that planets around giant stars move inward during the RGB and/or horizontal branch (HB) phase. 
Because of the effect of the tidal decay, planets around giant stars that reside between the survival limit and a maximum critical 
distance are expected to shrink their orbital radius. 
Therefore, as a result of the stellar evolution, these planets are expected to be absent in close-in orbits and on average closer 
than those around less evolved stars. 
On the other hand, planets that reside at much larger orbital distance avoid the strong tidal interaction and they increase their 
orbital radius due to the lower gravitational potential caused by the stellar mass loss.
This process thus might explain the difference in the orbital distance of planets
around giant stars and those orbiting subgiant stars that is observed in Figure \ref{semiaxis_star}. \newline \indent
This interpretation, however, has to be taken with caution. In particular, the uncertainty in the mass and 
evolutionary status of the host stars might be biasing this result. 
Lloyd (\cite{LLO11,LLO13}) claimed that the mass of the planet-hosting subgiants and first ascending giant stars
have been systematically overestimated or that many of such detections are actually false positives. If either of these is 
the actual case, then the previous argument is not valid, and hence the orbital
properties of planets around subgiant and giant stars cannot be directly compared because their host stars comes from two different 
populations (mainly low-mass and intermediate-mass stars, respectively). In such a case, it is not possible to disentangle the effect 
of the stellar mass and disk properties (migration mechanism, dissipation timescale, etc.) from the stellar evolution.
In addition, Bowler et al. (\cite{BOW10})
showed that the period-mass distribution of planets around solar-type and intermediate-mass stars are different at the 
4-$\sigma$ level. In particular, they showed that planets around subgiant stars with 1.5 \msun $<$ M$_\star$ $<$ 2.0 \msun
tend to reside at larger distances from the host star ($a$ > 1 AU) and that their occurrence rate is higher. 
Since the subgiants still have relatively small radii, this is not expected to be attributed to tidal decay, and hence it reveals
a primordial different formation and evolution of planets around more massive stars.
This result is incompatible with Lloyd's first alternative in the sense that if these subgiant stars are actually low-mass stars, then their
planets should exhibit similar properties to those exhibited by systems around solar-type stars.

\subsection{Mass distribution}

\begin{figure*}[ht]
\centering
\includegraphics[width=13cm,height=16cm,angle=270]{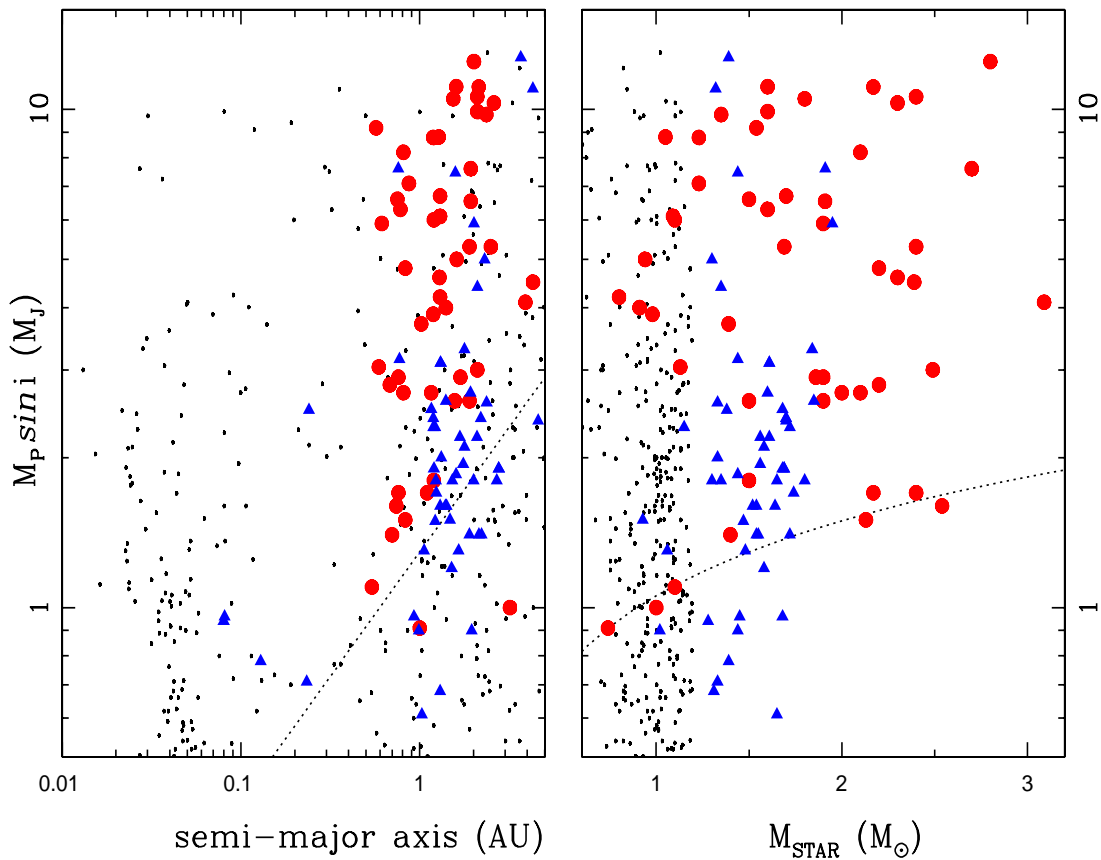}
\caption{Minimum mass of the planet versus the orbital distance and mass of the host star (left and right panel, respectively).
The blue triangles and red filled circles correspond to subgiant and giant
host stars, respectively. The small black dots show the position of known planets around solar-type stars.
The dashed lines correspond to a RV amplitude of 30 \ms, for a planet orbiting
a 1.5 \msun star (left panel) and to an orbital distance of 1 AU (right panel).
\label{semiaxis_star}}
\end{figure*}

\begin{figure}[h]
\centering
\includegraphics[width=7cm,height=9cm,angle=270]{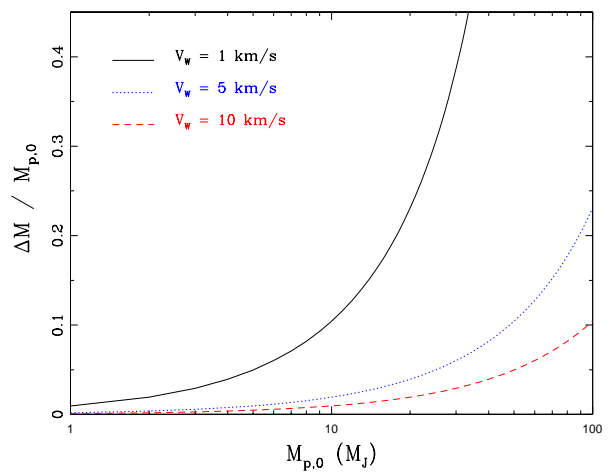}
\caption{Fractional amount of mass accreted by a planet as a function of its initial mass. The three horizontal lines correspond to
wind velocities at the planet's position of 1, 5 and 10 k\ms, respectively.
\label{accretion}}
\end{figure}

\begin{figure}[h]
\centering
\includegraphics[width=7cm,height=9cm,angle=270]{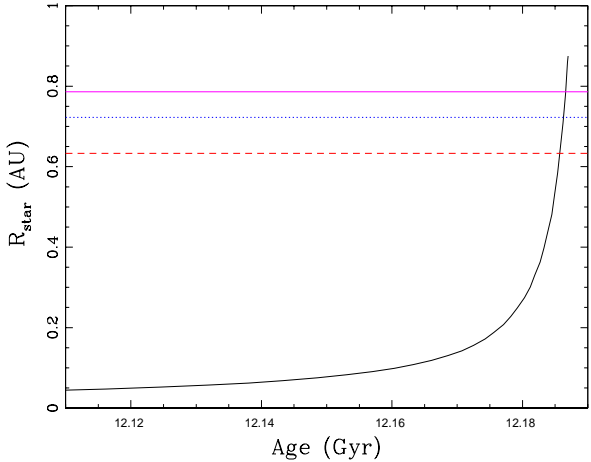}
\caption{Evolution of a 1 \msun star radius, at the end of the RGB phase (from Salasnich et al. \cite{SAL00} models).
The three horizontal lines (from bottom to top) show the Roche radius of the
star in the presence of a planet at 1 AU, with masses of 50, 10 and 1 \mjup, respectively.
\label{roche_lobe}}
\end{figure}

Although the current spectrographs, such as the High Accuracy Radial Velocity Planet Searcher (HARPS; Mayor et al. \cite{MAY03}), allow us to reach a RV precision 
at the sub-\ms level, the detection of planets around evolved stars via the RV technique is restricted to gas giants, 
since stellar oscillations make the detection of RV signals below the $\sim$ 30 \ms level more difficult.
However, it is still possible to study the properties of such massive planets, and compare them with the population of gas 
giants around solar-type stars. \newline \indent
The first indication of the peculiar mass distribution of planets around evolved stars was presented by Lovis \& Mayor 
(\cite{LOV07}) whom, based on a very restricted sample, suggested that there was an abnormally high 
fraction of massive planets and brown dwarfs around post-MS stars. Afterward, this result was confirmed by D\"{o}llinger et al. (\cite{DOL09}), 
who showed that the giant planets around giant stars are more common and
on average more massive than those detected in solar-type stars. This observational result is clearly observed in Figure
\ref{semiaxis_star}. As can be seen, most of the giant planets around giant stars are super planets (M$_p$ $\gtrsim$ 
3 \mjup). 
This is in stark contrast to the mass distribution of giant planets around low-mass stars, since 
most of them have masses below $\sim$ 2 \mjup (source:http://exoplanet.eu). This result can be interpreted to be caused 
by the influence of the host star mass on the properties of substellar companions. 
It seems natural to expect that more massive
stars have more massive and denser disks, from where massive planets are efficiently formed. However, it can be seen 
in Figure \ref{semiaxis_star} (right) that there is no dependence of the planet's mass with the stellar mass. Moreover,
the mass distribution of planets around giant stars is completely different than for planets orbiting subgiants.  
To investigate the difference between the two distributions, let us consider only those planets around stars
with 0.9 \msun $<$ M$_\star$ $<$ 2.0 \msun. A K-S test yields a null hypothesis probability of only 3\,$\times\,10^{-7}$,
meaning that the two datasets are drawn from different distributions. 
\newline \indent
Although it is difficult to interpret this result, it suggests that planets grow during the giant phase
of the host star. 
One possibility is that planets accrete material directly from the stellar envelope or from the stellar wind during the 
RGB phase.
These ideas have been studied from a theoretical point of view in the past. For instance, Livio \& Soker 
(\cite{LIV83},\cite{LIV84}) showed that planets with M$_p$ $\gtrsim$ 13 \mjup can survive inside the stellar envelope 
thus accreting a significant amount of mass. As a result, the planet ends up in a very close-in orbit and having a mass 
of $\sim$ 0.14 \msun. However , they also showed that smaller planets are either evaporated in the stellar envelope or they 
rapidly spiral inward and finally collide with the stellar core. 
\newline \indent
On the other hand, accretion from the stellar wind seems to be a plausible growing mechanism, since it is not expected to
affect the planet's orbit during this process considerably. Duncan \& Lissauer (\cite{DUN98}) showed that the total amount of 
material hitting the planet during the entire RGB phase is only a fraction of 
the planet's mass, meaning that this mechanism cannot solely be responsible for this observational result. 
Spiegel \& Madhusudhan (\cite{SPI12}) confirmed the previous results, but they also showed that through this mechanism the planet can 
accrete a significant amount of mass in its atmosphere, thus modifying its composition. 
We performed similar calculations, considering Bondi-Hoyle accretion (Bondi \& Hoyle \cite{BON44}), without planet evaporation. 
We integrated the stellar mass-loss formula given by (Reimers \cite{REI75}), assuming a spherical distribution, through the entire 
RGB phase, using a 1.0 \msun star model from Salasnich et al. (\cite{SAL00}).
We used a value of $\eta$ = 2.6\,$\times\,10^{-13}$, so that the total mass lost during the RGB is 0.33 \msun, as predicted by 
Schr\"oder \& Connon Smith (\cite{SCH08}).
The orbital distance of the planet was fixed to 1.0 AU. At each time step we updated the mass of the planet. 
Figure \ref{accretion} shows the normalized total accreted mass as a function of the initial mass of the planet. The three lines 
correspond to wind velocities at the planet's position of 1, 5 and 10 k\ms, respectively.
As can be seen, a giant planet (M$_{\rm p}$ $\lesssim$ 10 \mjup) is able to accrete a significant amount of mass only for very low-wind-velocity 
($\sim$ 1 k\ms). For higher wind velocities ($\gtrsim$ 5 k\ms), the total accreted mass is negligible. This result is also valid for planets around
intermediate-mass stars, since the mass-loss rate is even lower. However, the total accreted mass by the planet might be 
significantly higher in the case of non-spherical or enhanced mass-loss (e.g. Tout \& Eggleton \cite{TOU88}). 
If the star loses mass preferentially through 
a disk aligned with the planet's orbital plane, then the density of the wind at the planet's position is much higher. On the 
other hand, an enhanced mass-loss rate, driven for instance by the deposition of angular momentum in the stellar envelope by 
other planets in the system (Soker \cite{SOK98}), might lead to a much higher total accreted mass.
For more massive substellar 
objects (M $\gtrsim$ 50 \mjup), the accretion rate is much higher and thus a substellar companion might be able to accrete a
total of several Jupiter masses at wind velocities of 5-10 k\ms and at an orbital distance up to a few AU. Moreover, in the 
low-wind-velocity regime, these objects accrete enough material to become a low-mass star.
This result should be investigated in more detail in the future, since this mechanism can be responsible 
for the lack of massive brown dwarfs in relatively close-in orbits around post-MS stars. 
\newline \indent
A third possibility is that the star overfills its Roche lobe, thus directly transferring material to the planet. 
Figure \ref{roche_lobe} shows the evolution of the stellar radius (1 \msun model described above), at the end of the RGB phase.  
The solid, dotted and dashed lines correspond to the Roche radius (see Eggleton \cite{EGG83}) of the star in the presence of an orbiting 
planet at 1 AU with M$_{\rm P}$ = 1, 10 and 50 \mjup, respectively. The star overfills its Roche lobe
before reaching the tip of the RGB phase, meaning that there is material that is transferred directly from the stellar envelope to the 
planet. Also, as can be seen in Figure \ref{roche_lobe}, the radius of the star at this stage is comparable in size with the orbital 
distance, which also is accompanied by a strong tidal effect that would eventually lead to the planet engulfment.  
However, if the tidal decay is weak enough so that the planet avoids engulfment, the planet can accrete material from the wind and 
later on via direct mass transfer, reducing its orbital distance (because of the accretion process and by tidal decay) and increasing its 
mass. This scenario might explain the peculiar orbital and mass distributions revealed by planets orbiting giant stars.  \newline \indent
Finally, these planets might be the remnants of either brown dwarfs or low-mass stellar companions that 
lost their envelope during the common-envelope phase. The main problem with this idea is that this 
process leads to a significant loss of the stellar envelope, and thus to the formation of an extreme horizontal branch star 
(Soker \cite{SOK98}; Han et al. \cite{HAN02}). 

\subsection{Eccentricity}
\begin{figure}[h]
\centering
\includegraphics[width=9cm,height=9cm,angle=270]{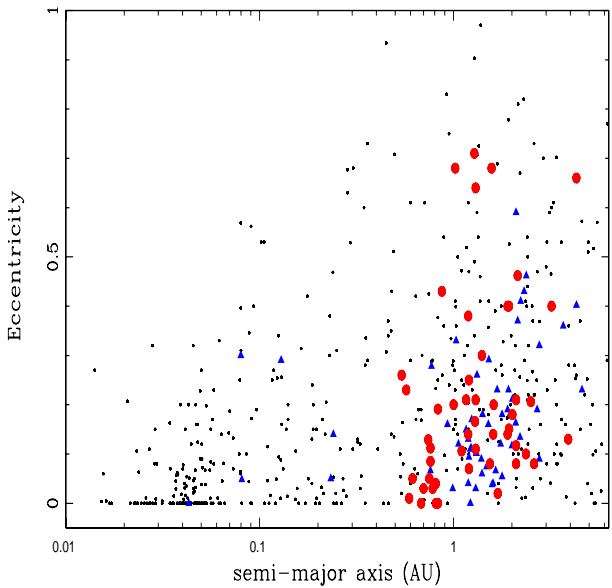}
\caption{Planet's eccentricity against the orbital distance.
The blue triangles and red filled circles correspond to subgiant and giant
host stars, respectively. The small black dots show the position of known planets around solar-type stars.
\label{semiaxis_eccentricity}}
\end{figure}

One of the most surprising properties of exoplanets is their relatively large orbital eccentricity.
Figure \ref{semiaxis_eccentricity} shows the eccentricity as a function of planets semi-major axis.
The symbols are the same as in Figure \ref{semiaxis_star}.
As can be seen in Figure \ref{semiaxis_eccentricity}, there are two distinctive populations: planets orbiting interior
to $\sim$ 0.1 AU, which present relatively low eccentricities, and those at larger orbital distances, where there are many highly 
eccentric systems.  \newline \indent 
The origin of the large eccentricities observed in extrasolar planets is puzzling, since the eccentricity is predicted to
be efficiently damped during the early planet formation stage via interaction of the planet with the gaseous disk (e.g. Artymowicz 
\cite{ART93}; Tanaka \& Ward \cite{TAN04}). As a result of this process, planets are expected to have nearly circular orbits after 
their formation. Different mechanisms might be responsible for pumping-up the orbital eccentricity, such as 
planet-planet interactions (e.g. Lin \& Ida \cite{LIN97}; Marzari \& Weidenschilling \cite{MAR02}) and planet-star interactions (e.g. 
Holman et al. \cite{HOL97}; Zakamska \& Tremaine \cite{ZAK04}). 
In this context, the low $e$ values exhibited by the short-period systems are likely explained by the subsequent effect of tidal 
circularization (e.g. Goldreich \& Soter \cite{GOL66}; Rasio et al. \cite{RAS96}; Jackson et al. \cite{JAC08}), similar to what 
is seen in short period binary stars (Duquennoy \& Mayor \cite{DUQ91}; Verbunt \& Phinney \cite{VER95}). \newline \indent
Additionally, it seems that planets around post-MS stars have lower eccentricities than those orbiting MS stars. 
Let us consider the region between $a$ $\sim$ 0.5\,-\,3.0 AU, where most of the planets around evolved stars reside. 
While 32 \% of them have $e$ $\leq$ 0.1, only 19\% of the planets around solar-type stars 
exhibit such low eccentricities. 
Moreover, considering $e$ up to 0.2, we obtain a fraction of 70\% for planets around subgiants, 64\% for planets orbiting giants and 38\% for 
system hosted by solar-type stars.  
These differences are significant and point out a different eccentricity distribution between planets around solar-type and
post-MS stars. In fact, considering only the exoplanets orbiting in the above mentioned region, a K-S test gives a probability of 
5.1$\times$10$^{-4}$ that the eccentricities of planets around giant stars and solar-type stars are drawn from the same 
distribution.  Similarly, when the subgiants and solar-type host stars are compared, the K-S test yields a probability of 
2.1$\times$10$^{-4}$. \newline \indent
This observational fact might be interpreted as the result of tidal circularization, especially for giants stars, which have 
much larger radii than solar-type stars. In fact, although previous works included only the effect of tides in the planets, 
neglecting the influence of tides in the star, Jackson et al. (\cite{JAC08}) showed that this 
term, which is strongly dependent on the stellar radius, might reduce the circularization timescale considerably. In this
scenario however, the planets around subgiant stars should present larger eccentricities, because of their still relatively small radii
and the short timescales involved. \newline \indent
On the other hand, this result might be telling us that a significant fraction of the MS star systems are actually multiplanets 
systems with circular orbits, which are misinterpreted as eccentric single planets (Anglada-Escude et al. \cite{ANG10}; Wittenmyer
et al. \cite{WIT12}), especially when they are in resonant orbits (e.g. Jenkins et al. \cite{JEN13b}). \newline \indent 
Last, since most of the planet-hosting evolved stars are more massive than solar-type stars, the low $e$ values observed 
in planets around these stars might be explained by a different formation scenario. For instance, if multiplanets systems are 
not efficiently formed in disks surrounding more massive stars, then the gravitational perturbation between planets is strongly
reduced. As a consequence, no significant enhancement of the planet's eccentricity by planet-planet interactions would be possible.

\section{Summary and conclusions}

We have reported the discovery of a substellar object in the planetary mass regime around the RGB star
HIP\,105854. The main orbital properties of the planet are the following: M$_{\rm p}$\,sin$i$ = 8.2 $\pm$ 0.2 \mjup; P = 184.2 $\pm$ 0.5 days; 
e = 0.02 $\pm$ 0.03.  This is the second planet detected during our survey and adds up to a growing 
population of substellar companions that have been detected around evolved stars. \newline \indent
Based on the properties of more than 100 planets around post-MS stars, we have discussed the effect 
of the host star evolution on the orbital and mass distributions of such objects. We have shown that the planets around 
giant stars tend to reside at smaller orbital distances than those around subgiant stars. While most of the planets around
subgiant stars have semi-major axes greater than $\sim$ 1 AU, there is an overabundance of planets around giant stars orbiting
between $\sim$ 0.5-0.9 AU. This observational result might be explained in part due to the loss of angular momentum of planets
during the RGB phase. In addition, short period planets are expected to be engulfed by the host star during the RGB phase, 
since the radius of the star becomes larger than the orbital distance at the end of the RGB phase, and that is why they are 
present only around subgiant stars and not around giant stars. \newline \indent
We also showed that the minimum mass distributions of planets around subgiant and giant stars are completely different, and that 
there is no dependence on the mass of the host star. This observational result suggests that there is an 
evolutionary effect and thus the planets grow during the RGB or HB phase. 
We proposed different mechanisms such as accretion from the stellar wind and mass transfer after the host star overfills its Roche lobe,
which might explain some of the observational properties of these systems. Other possibilities, such a planet mergers, should be also 
considered. \newline \indent 
Finally, we studied the eccentricity of planets around post-MS stars. We showed that on average they present lower eccentricity 
when compared to solar-type host stars. The explanation of this observational fact is uncertain, particularly because no significant
difference is observed between planets around subgiant and giant stars. Probably the formation and evolution of planetary
systems that are formed around more massive stars depart from the systems orbiting low-mass stars.

\begin{acknowledgements}
M.J. acknowledges financial support by Fondecyt Projects \#3140607 and \#1120299.
P.R. acknowledges financial support from Fondecyt through grant \#1120299 and Conicyt-PIA Anillo ACT1120.
J.J. acknowledges funding by Fondecyt through grant 3110004, 
the GEMINI-CONICYT FUND and from the Comit\'e Mixto ESO-GOBIERNO DE CHILE. 
M.J., J.J, and P.R also acknowledge support from BASAL PFB-06 (CATA).
\end{acknowledgements}


\begin{thebibliography}{}
\bibitem[2010]{ANG10} Anglada-Escude, G, Lopez-Morales, M \& Chambers, J. E. 2010, \apj, 709, 168
\bibitem[1993]{ART93} Artymowicz, P. 1993, \apj, 419, 166
\bibitem[1944]{BON44} Bondi, H. \& Hoyle, F. 1944, MNRAS, 104, 273
\bibitem[1997]{BOS97} Boss, A. 1997, Science, 276, 1836
\bibitem[2010]{BOW10} Bowler, B. P., Johnson, J. A., Marcy, G. W. et al. 2010, \aap, 709, 396 
\bibitem[1996]{BUT96} Butler, R. P. et al. 1996, PASP, 108, 500
\bibitem[2004]{DEM04} Demarque, P., Woo, J., Kim, Y. \& Yi, S. K. 2004, \apjs, 155, 667
\bibitem[2009]{DOL09} D\"ollinger, M. P., Hatzes, A. P., Pasquini, L., Guenther, E. W. \& Hartmann, M. 2009, \aap, 505, 1311
\bibitem[2008]{DUN98} Duncan M. J. \& Lissauer, J. J. 1998, Icarus, 134, 303
\bibitem[1991]{DUQ91} Duquennoy, A. \& Mayor, M. 1991, \aap, 248, 485 
\bibitem[1983]{EGG83} Eggeleton, P. P. 1983, \apj, 268, 368 
\bibitem[2005]{FIS05} Fischer, D. A. \& Valenti, J. 2005, \apj, 622, 1102
\bibitem[2001]{FRI01} Frink, S., Quirrenbach, A., Fischer, D.,  R\"oser, S. \& Schilbach, E. 2001, PASP, 113, 173
\bibitem[1966]{GOL66} Goldreich, P. \& Soter, S. 1966, Icarus, 5, 375
\bibitem[1997]{GON97} Gonzalez, G. 1997, \mnras, 285, 403
\bibitem[2002]{HAN02} Han, Z., Podsiadlowski, Ph., Maxted, P. F. L., Marsh, T. R. \& Ivanova, N. 2002, \mnras, 336, 449
\bibitem[2002]{HAT02} Hatzes, A. P. 2002, AN, 323, 392
\bibitem[1997]{HOL97} Holman, M., Touma, J. \&  Tremaine, S. 1997, Nature, 386, 254
\bibitem[2008]{JAC08} Jackson, B., Greenberg, R. \& Barnes, R. 2008, \apj, 678, 1396
\bibitem[2008]{JEN08} Jenkins, J. S. et al. 2008, \aap, 485, 571
\bibitem[2011]{JEN11} Jenkins, J. S. et al. 2011, \aap, 531, 8
\bibitem[2013a]{JEN13a} Jenkins, J. S., Jones, H. R. A., Tuomi, M. et al. 2013a, \apj, 766, 67
\bibitem[2013b]{JEN13b} Jenkins, J. S, Tuomi, M., Brasser, R., Ivanyuk, O. \& Murgas, F. 2013b, \apj, 771, 41 
\bibitem[2007]{JOH07} Johnson, J. A., Fischer, D. A., Marcy, G. W. et al. 2007, A\&A, 665, 785 
\bibitem[2011]{JON11} Jones, M. I., Jenkins, J. S., Rojo, P. \& Melo, C. H. F. 2011, A\&A, 536, 71 
\bibitem[2013]{JONES13} Jones, M. I. 2013, PhD Thesis, Universidad de Chile
\bibitem[2013]{JON13} Jones, M. I., Jenkins, J. S., Rojo, P., Melo, C. H. F. \& Bluhm, P. 2013, A\&A, 556, 78
\bibitem[2014]{JONJEN14} Jones, M. I. \& Jenkins, J. S. 2014, A\&A, 562, 129 
\bibitem[1999]{KAU99} Kaufer, A., Stahl, O., Tubbesing, S. et al. 1999, The Messenger 95, 8
\bibitem[2008]{KEN08} Kennedy, G. M. \& Kenyon, S. J. 2008, \apj, 673, 502
\bibitem[2011]{KUN11} Kunitomo, M., Ikoma, M., Sato, B. et al. 2011, \apj, 737, 66 
\bibitem[2014]{LILLO14} Lillo-Box, J., Barrado, D., Moya, A. et al. 2014, A\&A, 562, 109
\bibitem[1997]{LIN97} Lin, D. N. C. \& Ida, S. 1997, \apj, 477, 781
\bibitem[1983]{LIV83} Livio, M. \& Soker, N. 1983, \aj, 125, 12
\bibitem[1984]{LIV84} Livio, M. \& Soker, N. 1984, \mnras, 208, 763
\bibitem[2007]{LOV07} Lovis, C. \& Mayor, M. 2007, \aap, 472, 657
\bibitem[2011]{LLO11} Lloyd, J. P. 2011, \apj, 739, 49
\bibitem[2013]{LLO13} Lloyd, J. P. 2013, \apj, 774, 2    
\bibitem[2002]{MAR02} Marzari, F. \& Weidenschilling, S. J. 2002, Icarus, 156, 570
\bibitem[2008]{MAS08} Massarotti, A., Latham, D., Stefanik, R. P. \& Fogel, J. 2008, \aj, 135, 209 
\bibitem[2003]{MAY03} Mayor, M., Pepe, F., Bouchy, F, et al. 2003, The Messenger 114, 20
\bibitem[2009]{MES09} Meschiari, S., Wolf, A. S., Rivera, E. et al. 2009, PASP, 121, 1016 
\bibitem[2013]{MOR13} Mortier, A., Santos, N. C., Sousa, S. G. et al. 2013, A\&A, 557, 70
\bibitem[2007]{PAS07} Pasquini, L., D\"ollinger, M. P., Weiss, A. et al. 2007, \aap, 473, 979
\bibitem[1999]{QUE99} Queloz, D., Casse, M. \& Mayor, M. 1999, ASPC 185, 13
\bibitem[1999]{RAN99} Randich, S., Gratton, R., Pallavicini, R., Pasquini, L. \& Carretta, E. 1999, \aap, 348, 487 
\bibitem[1996]{RAS96} Rasio, F. A., Tout, C. A., Lubow, S. H. \& Livio, M. 1996, \apj, 470, 1187
\bibitem[1975]{REI75} Reimers, D. 1975, Mem. Soc. R. Sci. Liege, 6th Ser. 8, 369
\bibitem[2000]{SAL00} Salasnich, B., Girardi, L., Weiss, A. \& Chiosi, C. 2000, A\&A, 361, 1023 
\bibitem[2001]{SAN01} Santos, N. C., Israelian, G. \& Mayor, M. 2001, \aap, 373, 1019
\bibitem[2005]{SAT05} Sato, B., Kambe, E., Takeda, Y. et al. 2005, \pasj, 57, 97 
\bibitem[2008]{SAT08} Sato, B., Izumiura, H., Toyota, E. et al. 2008, \pasj, 60, 539 
\bibitem[1982]{SCA82} Scargle, J. D. 1982, \apj, 263, 835 
\bibitem[2008]{SCH08} Schr\"oder, K.-P. \& Connon Smith, R. 2008, MNRAS, 386, 155
\bibitem[2005]{SCH05} Schuler, S. C., Kim, J. H., Tinker, M. C. \& King, J. R. 2005, \apj, 632, 131
\bibitem[1998]{SOK98} Soker, N. 1998, \aj, 116, 1308
\bibitem[2012]{SPI12} Spiegel, D. S. \& Madhusudhan N. 2012, \apj, 756, 132
\bibitem[2004]{TAN04} Tanaka, H. \& Ward, W. R. 2004, \apj, 602 388
\bibitem[2013]{TOK13} Tokovinin, A., Fischer, D. A., Bonati, M. et al. 2013, PASP, 125, 1336
\bibitem[1988]{TOU88} Tout, C. A. \& Eggleton, P. P. 1988, \mnras, 231, 823
\bibitem[1995]{VER95} Verbunt, F. \& Phinney, E. S. 1995, \aap, 709, 721
\bibitem[2009]{VIL09} Villaver, E. \& Livio, M. 2009, \apj, 705, 81 
\bibitem[2012]{WIT12} Wittenmyer, R. A.,Horner, J., Tuomi, M. et al. 2012, \apj, 753, 169
\bibitem[2004]{ZAK04} Zakamska, N. L. \& Tremaine, S. 2004, \aj, 128, 869
\end{thebibliography}
\end{document}